\input{psfig}

\documentstyle[11pt,aaspp4,epsf]{article}

\def \ltsima{$\; \buildrel < \over \sim \;$}
\def \simlt{\lower.5ex\hbox{\ltsima}}            
\def \gtsima{$\; \buildrel > \over \sim \;$}
\def \gtsima{\mbox{$\; \buildrel > \over \sim \;$}}
\def \simgt{\lower.5ex\hbox{\gtsima}} 
\newcommand{\etal}{{et al.}~}
\newcommand{\eg}{{e.g.~}}

\begin{document}

\pagestyle{plain}
\setcounter{page} {1}
\begin{center}
{\Large \bf Evolution of the cosmological mass function in a moving barrier model}
\end{center}
\vskip2cm
\begin{center}
{\bf Antonino Del Popolo$^1$,$^2$}
\\~
\\~
$1$  Bo$\breve{g}azi$\c{c}i University, Physics Department,
     80815 Bebek, Istanbul, Turkey\\
$^2$ Dipartimento di Matematica, Universit\`{a} Statale di Bergamo, Piazza Rosate 2, Bergamo, ITALY


\end{center}

\vskip 2cm

\begin{abstract}
I calculate the mass function evolution in a $\Lambda$CDM model by means of the excursion set model and an improved version of the barrier  
shape obtained in Del Popolo \& Gambera (1998), which implicitly takes account of the total angular momentum acquired by the proto-structure during evolution and of a non-zero cosmological constant. I compare the result with Reed et al. (2003), who used a high resolution $\Lambda$CDM numerical simulation to calculate the mass function
of dark matter haloes down to the scale of dwarf galaxies, back to a redshift of
fifteen.
I show that the mass function obtained in the present paper, gives similar predictions to the Sheth \& Tormen mass function but it does not show the
overprediction of extremely rare objects shown by the Sheth and Tormen mass function.
The results confirm previous findings that the simulated halo mass function can be described solely by the variance of the mass distribution, 
and thus has no explicit redshift dependence.

%

\end{abstract}

\keywords{cosmology: theory - large scale structure of Universe - galaxies: formation}



\section{Introduction}

The universe that we observe appears quite clumpy and inhomogeneous on a spatial scale of $ \simeq 200 h^{-1}$ Mpc. Beyond that scale mass clumps appear to be homogeneously distributed. We observe massive clumps such as galaxies, groups of galaxies, clusters of galaxies, super-clusters (the most massive one among the hierarchy of structures) fill the space spanning a wide range of mass scales. All these objects, together, form the structure that is known as the Large Scale Structure (LSS) in the universe. One of the most fundamental challenges in the present universe is to understand the formation and evolution of the LSS. In order to understand the LSS, we need to have a theoretical framework within which predictions for structure formation can be made. The leading idea of all structure formation theories is that structures
were born from the collapse of small Gaussian density fluctuations originated from quantum fluctuations
(Guth \& Pi 1982; Hawking 1982; Starobinsky 1982; Bardeen et al. 1986 -  hereafter BBKS).
Starting from these fluctuations, collapsed, virialized dark matter haloes condensed out. Within these haloes
gas cools and stars form (White \& Reese 1977; White \& Frenk 1991; Kauffmann et al. 1999). As a consequence, the structure of dark matter haloes is of fundamental importance in the
study of the formation and evolution of galaxies and
clusters of galaxies. From the theoretical point of view, the structure of
dark matter haloes can be studied both analytically and numerically.
An analytical model that has achieved a wide popularity is the Press-Schecter (1974) (hereafter PS) formula, which 
together with its extensions (EPS) are of great interest since they allow us
to compute mass functions (Press \& Schechter  1974; Bond et al.
1991), to approximate merging histories (Lacey \& Cole 1993, LC93
hereafter, Bower 1991, Sheth \& Lemson 1999b) and to estimate the
spatial clustering of dark matter haloes (Mo \& White 1996;
Catelan et al. 1998, Sheth \& Lemson 1999a).\\

Although the analytical framework of the PS model has been greatly refined and extended (as testified by the previous 
cited papers), it is well known that the PS mass function, while qualitatively correct, disagrees with the results of 
N-body simulations. Ellipsoidal halo collapse models (e.g. Monaco 1997ab; Lee \& Shandarin 1998; Sheth, Mo,
\& Tormen 2001 (hereafter SMT)) yield much more robust predictions than the
conventional spherical collapse models, and are in excellent agreement
with empirical fits by Sheth \& Tormen (1999) (hereafter ST) or Sheth \& Tormen (2002) (hereafter ST1).  Monaco et al. (2002),
using the semi-analytic code PINOCCHIO, which uses a perturbative
approach, show that the dark matter halo distribution can be
accurately predicted at much lower computational cost, on a
point-by-point basis, from a numerical realization of an initial
density field.  Jenkins et al. (2001) (hereafter J01) utilized a large set of
simulations of a range of volumes and cosmologies (Jenkins et al.
1998; Governato et al. 1999; Evrard et al. 2002) to test the ST mass function over more than four orders of magnitude in
mass, and out to a redshift of 5, finding good agreement with the ST
function down to their resolution limit of $\simeq 3 \times 10^{11} M_{\odot}$, except for an overprediction by the ST function for haloes
at rare density enhancements. Reed et al (2003) (hereafter R03) used a high resolution $\Lambda$CDM numerical simulation to calculate the mass function
of dark matter haloes down to the scale of dwarf galaxies, back to a redshift of
fifteen, in a 50 $h^{-1}$Mpc volume containing 80 million particles.  
They probed previously untested regimes of the mass function by simulating a
volume that resolves haloes down to the scale of $10^{10} M_{\odot}$ 
dwarfs, in a cosmological environment, allowing us to sample the mass function back
to z$\simeq$15.
They considered the possibility of an empirical
adjustment to the ST function.  More recently, Yahagi, Nagashima \& Yoshii (2004) (hereafter YNY) 
performed five runs of N-body simulations
with high mass resolution and compared them with different multiplicity function and with a fit by them proposed. 
They showed that discrepancies are observed  
between some of the quoted analytical multiplicity function with high mass resolution simulations:
for example the maximum value of the multiplicity function from their simulations at $\nu \simeq 1$ 
(where $\nu=\left(\frac{\delta_{\rm co}(t)}{\sigma(M)}\right)^2$, being $\sigma^2(M)$ the present day mass dispersion on comoving scale
containing mass $M$, and $\delta _{\rm co}=1.68$ is the critical threshold for a spherical model)
is smaller, and its low
mass tail is shallower when compared with the ST, ST1 multiplicity function. Warren et al. (2005) (W05) determined the mass
function, as well as its uncertainty, using sixteen $1024^3$-particle nested-volume
dark-matter simulations, spanning a mass range of over five orders of magnitude. 
They also considered the possibility of an empirical
adjustment to the ST function.  

In the present paper, I use an improvement of the barrier shape given in Del Popolo \& Gambera (1998), obtained from the parameterization of the nonlinear collapse discussed in that paper, together with the results of ST1 in order to study the evolution of the mass function. 
I show that the function obtained in the present paper provides a better fit than the ST or other functional forms used in literature to R03 simulations data and moreover that it has been obtained from solid physical, 
theoretical, arguments.

The paper is organized as follows: in Sect. ~2, I calculate the mass function.
In Sect. ~3, I discuss the results and 
Sect. ~4 is devoted to conclusions.

\section{The mass function and the moving barrier}

According to hierarchical scenarios of structure formation, a region collapses
at time $t$ if its overdensity at that time exceeds some threshold. The linear
extrapolation of this threshold up to the present time is called a barrier, $B$.
ST and ST1 provided formulas to calculate these last quantities starting from the shape of the 
barrier. In the following, I'll use an improved version of the barrier obtained in Del Popolo \& Gambera (1998) to get the mass function.

In order to calculate the barrier shape, it is possible to follow Del Popolo \& Gambera (1998) model, summarized in the following.
Let's consider an ensemble of gravitationally growing mass concentrations 
and suppose that the material in each system collects within the
same potential well with inward pointing acceleration given by $g(r)$ (see Del Popolo \& Gambera 1998). I
indicate with $dP=f(L,r v_r,t)dL dv_r dr$ the probability that a particle of mass $m$ 
can be found  in the proper radius range $r$, $r+dr$, in the radial
velocity range $v_r={\dot r}$, $v_r+d v_r$ and with angular momentum
$L= m r v_\theta$ in the range $dL$, or specific angular momentum $l=L/m=r v_\theta$, where $v_\theta$ is the tangential velocity.
%
Assuming a non-zero cosmological constant, 
the radial
acceleration of the particle is: 
\begin{equation}
\frac{dv_r}{dt}=-\frac{G M }{r^2}+\frac{l^2(r)}{r^3}+\frac{\Lambda}{3} r \label{eq:coll}
\end{equation}
(Peebles 1993; Bartlett \& Silk 1993; Lahav 1991; Del Popolo \& Gambera 1998, 1999)
where $M$ is the mass of the central concentration.

Integrating Eq. (\ref{eq:coll}) we have: 
\begin{equation}
\frac{1}{2}\left( \frac{dr}{dt}\right) ^{2}=\frac{GM}{r}+\int
\frac{l^{2}}{r^{3}}dr+\frac{\Lambda }{6}r^{2}+\epsilon
\label{eq:coll1}
\end{equation}
where the value of the specific binding energy of the shell, $\epsilon$, can be obtained using the condition for turn-around, $\frac{dr}{dt}=0$.

In turn the binding energy of a growing mode solution is uniquely given
by the linear overdensity, $\delta_{i}$, at time $t_{i}$.
From this overdensity, using the linear theory, one may obtain that of
the turn-around epoch and then that of the collapse.
I find the binding energy of the shell, $C$, using the
relation between $v$ and $\delta_{i}$ for the growing mode
(Peebles 1980) in Eq. (\ref{eq:coll1}) and finally I find that the
linear overdensity at the time of collapse is given by:
\begin{equation}
B(M)=\delta _{\rm c} (\nu, z)=\delta _{\rm co}\left[ 1+
\int_{0}^{r_{\rm ta}}  \frac{r_{\rm ta} l^2 \cdot {\rm d}r}{G M r^3}+\Lambda \frac{r_{\rm ta} r^2}{6 G M}
\right] \simeq \delta _{\rm co} \left[
1+\frac{\beta_1}{\nu^{\alpha_1}}+\frac{\Omega_{\Lambda} \beta_2}{\nu^{\alpha_2}}
\right]
\label{eq:maa}
\end{equation}
where $\alpha_1=0.585$, $\beta_1=0.46$, $\alpha_2=0.4$ and $\beta_2=0.02$, 
$r_{\rm i}$ is the initial radius, $r_{\rm ta}$ is the turn-around radius, and $l$ the specific angular momentum. 
The specific angular momentum appearing in Eq. ~(\ref{eq:maa}) is directly proportional to the total angular momentum acquired by the proto-structure during evolution. In order to calculate $l$, it is possible to use the same model
as described in Del Popolo \& Gambera (1998, 1999) (more details on
the model and some of the model limits can be found in
Del Popolo, Ercan \& Xia 2001).

The CDM spectrum used to calculate the mass function is that of BBKS (equation~(G3)), with transfer function:
\begin{equation}
T(k) = \frac{[\ln \left( 1+2.34 q\right)]}{2.34 q}
\cdot [1+3.89q+
(16.1 q)^2+(5.46 q)^3+(6.71)^4]^{-1/4}
%
%
\label{eq:ma5}
\end{equation}
(where 
$q=\frac{k\theta^{1/2}}{\Omega_{\rm X} h^2 {\rm Mpc^{-1}}}$.
Here $\theta=\rho_{\rm er}/(1.68 \rho_{\rm \gamma})$
represents the ratio of the energy density in relativistic particles to
that in photons ($\theta=1$ corresponds to photons and three flavors of
relativistic neutrinos).
The power spectrum was normalized to reproduce the observed abundance of rich 
cluster of galaxies (e.g., Bahcal \& Fan 1998).

ST1 connected the form of the barrier with the form of the
mass function. 
As shown by ST1, for a given barrier shape, $B(S)$, 
the first crossing distribution is well approximated by:
\begin{equation}
f(S)dS=|T(S)|\exp (-\frac{B(S)^{2}}{2S})\frac{dS/S}{\sqrt{2\pi S}}
\label{eq:distrib}
\end{equation}   
where $S\equiv \sigma^2(M)$ and $T(S)$ is the sum of the first few terms in the Taylor expansion of $B(S)$:
\begin{equation}
T(S)=\sum_{n=0}^{5}\frac{(-S)^{n}}{n!}\frac{\partial ^{n}B(S)}{\partial S^{n}}
\label{eq:expans}
\end{equation}
In the case of the barrier shape given in Eq. (\ref{eq:maa}) 
the Eqs. ~(\ref{eq:distrib}),(\ref{eq:expans}), give, after truncating the expansion at $n=5$ (see ST),
the multiplicity function, $\nu f(\nu)$: 
\begin{equation}
\nu f(\nu )=A _1 \left( 1+\frac{\beta_1 g(\alpha_1)}{\left( a\nu \right) ^{\alpha_1}}
+\frac{\beta_2 g(\alpha_2)}{\left( a\nu \right) ^{\alpha_2}}
\right) \sqrt{\frac{a\nu }{2\pi }}\exp{\{-a \nu \left[ 1+\frac{\beta_1}{\left( a\nu \right) ^{\alpha_1}}
+\frac{\beta_2}{\left( a\nu \right) ^{\alpha_2}}
\right] ^{2}/2\}}
\label{eq:mia}
\end{equation}
\begin{equation}
g(\alpha_i)=
\mid 1-\alpha_i +\frac{\alpha_i (\alpha_i
-1)}{2!}-...-\frac{\alpha_i(\alpha_i-1)\cdot \cdot \cdot
(\alpha_i-4)}{5!} \mid
\end{equation}
where $A_1=1.75$, $i=1$ or 2, $\alpha_1=0.585$, $\beta_1=0.46$, $\alpha_2=0.4$ and $\beta_2=0.02$, $a=0.707$.

The ``multiplicity function" is correlated with the usual, more straightforwardly used, ``mass function" as follows.  
Following Sheth \& Tormen (1999) notation, if $f(m,\delta) dm$ denotes the fraction of mass that is contained in collapsed haloes that have mass in the range $m$-$m+dm$, at redshift $z$, and $\delta(z)$ the redshift dependent overdensity, 
the associated ``unconditional" mass function is:
\begin{equation}
n(m,\delta)dm=\frac{\overline{\rho}}{m} f(m,\delta) dm
\label{eq:mfu}
\end{equation}

In the excursion set approach, the universal or ``unconditional" mass function, $n(m,z)$, representing 
the average comoving number density of haloes of mass $m$ 
is given by:
\begin{equation}
n(m,z)=\frac{\overline{\rho(z)}}{m^{2}}\frac{d\log{\nu }}{d\log m}\nu f(\nu )
\label{eq:universall}
\end{equation}
(Bond et al. 1991), where $\overline{\rho}$ is the background density.
The function $\nu f(\nu)$ is the ``multiplicity function"
which is obtained by computing the distribution of first crossings, $f(\nu) d \nu$, of a barrier $B(\nu)$, by independent, uncorrelated Brownian motion random walks. 
Multiplicity function and mass function are related by Eq. (\ref{eq:universall}). It is to be noted that in literature sometime the terms mass function and multiplicity function are used as synonymous (e.g. ST, Lin et al. 2002).

In the case of the barrier with non-zero cosmological constant, 
a good approximation to the multiplicity function is given by:
\begin{equation}
\nu f(\nu ) \simeq A _2 \left( 1+\frac{0.1218}{\left( a\nu \right) ^{0.585}}
+\frac{0.0079}{\left( a\nu \right) ^{0.4}}
\right) \sqrt{\frac{a\nu }{2\pi }}\exp{\{-0.4019 a \nu \left[ 1+\frac{0.5526}{\left( a\nu \right) ^{0.585}}
+\frac{0.02}{\left( a\nu \right) ^{0.4}}
\right] ^{2}\}}
\label{eq:mia1}
\end{equation}
where $A_2=1.75$.

With a similar calculation, ST1 found that  
\begin{equation}
\nu f(\nu)\simeq A_3 \left( 1+\frac{0.094}{\left( a\nu \right) ^{0.6}}\right) \sqrt{\frac{a\nu }{2\pi }}\exp{\{-a\nu \left[ 1+\frac{0.5}{\left( a\nu \right) ^{0.6}}\right] ^{2}/2\}}
\label{eq:sstt1}
\end{equation}
with $A_3 \simeq 1$.
This last result is in good agreement with the fit of the simulated first crossing distribution (ST):
\begin{equation}
\nu f(\nu )d\nu =A_4 \left( 1+\frac{1}{\left( a\nu \right) ^{p}}\right) \sqrt{\frac{a\nu }{2\pi }}\exp (-a\nu /2)
\label{eq:ssttt}
\end{equation}
where $p=0.3$, and $a=0.707$. 

The normalization factor $A_4$ has to satisfy the constraint:
\begin{equation}
\int_0^{\infty} f(\nu) d \nu=1
\end{equation}
and as a consequence it is not an independent parameter, but is expressed in the form:
\begin{equation}
A_4=\left[1+2^{-p} \pi^{-1/2} \Gamma(1/2-p)\right]^{-1}=0.3222
\end{equation}

R03 used a high resolution $\Lambda$CDM numerical simulation to calculate the mass function
of dark matter haloes down to the scale of dwarf galaxies, back to a redshift of
fifteen, in a 50 $h^{-1}$Mpc volume containing 80 million particles.  
Their low redshift results allow us to probe low $\sigma$ density fluctuations
significantly beyond the range of previous cosmological simulations.
They also 
considered the possibility of an empirical adjustment to the ST function. They inserted a crude multiplicative
factor to the ST function 
as follows, with $\delta_{\rm co}$ $=$ 1.686 and FOF $ll=$0.2
\footnote{
\bf 
The {\it friends-of-friends} (FOF) algorithm (Davis et al. 1985), is a halo finder algorithm which uses a ``linking length", ll, to link
together all neighboring particles with spacing closer than ll as members of a halo.
In R03 they utilize the FOF algorithm for the bulk of their analyses 
since it is reasonably robust and computationally efficient. 
Their FOF ll choice is 0.2 for all redshifts; 
this approach has been shown to be sound for a range of cosmologies by Jenkins
et al. (2001), although the evolution of parameters in the spherical collapse tophat model implies that ll should range from
ll =0.164 at z=0 to ll =0.2 at high redshift in CDM cosmology (Lacey \& Cole 1994; Eke, Cole, \& Frenk 1996; Jenkins et al. 2001). 
A deeper discussion can be found in R03 Sect. 4.1.
}:
\begin{equation}
f(\sigma) = f(\sigma; {\rm S{\rm}T})\bigg[exp[-0.7/(\sigma
[\cosh(2\sigma)]^5)]\bigg],
\label{eq:reed}
\end{equation}
valid over the range of -1.7 $\leq
\ln\sigma^{-1} \leq$ 0.9.  The resulting function
is virtually identical to the ST function
for all $-\infty \leq  \ln\sigma\leq 0.4$. At higher values
of $\ln\sigma^{-1}$, this function declines relative to the ST
function, reflecting an under abundance of haloes that becomes greater
with increasing $\ln\sigma^{-1}$.  For -1.7 $\leq \ln\sigma^{-1} \leq$
0.5, Eq. (\ref{eq:reed}) matches R03 data to better than 10$\%$ for well-sampled
bins, while  for 0.5 $\leq \ln\sigma^{-1} \leq$ 0.9, where Poisson
errors are larger, data is matched  to roughly 20$\%$. 
Note that in their notation, similarly to J01:
\begin{equation}\label{deff}
f(\sigma, z) \equiv \frac{M}{\overline{\rho(z)}}{{\rm d}n(M, z)\over{\rm
d}\ln\sigma^{-1}}
\end{equation}
Using the relation $\nu=(\frac{\delta_{\rm co}}{\sigma})^2$ and Eq. (\ref{eq:universall}) one finds that:
\begin{equation}
f(\sigma,z)=2 \nu f(\nu)
\end{equation}
%
%

The evolution of the mass function can be calculated evaluating it at different redshifts. The mass function, Eq. (\ref{eq:universall}), depends on redshift because of the dependence of $\overline{\rho(z)}$, $\nu$ or $\sigma$ from $z$. 
For example $\sigma(M,z) = \sigma(M,z=0)b(z)$, where ${\it b(z)}$ evolves as
$(1+z)^{-1}$ in an $\Omega_0 = 1$ universe, and more slowly in a
$\Lambda$CDM universe.  In the following I shall calculate the evolution of the mass function of the present paper and I compare it with R03 results.

\section{Results}

In this section, I compare the analytic mass function of the present paper with that of ST, and with R03 simulation results at several
redshifts. The PS mass function is not compared because, as is well known, while qualitatively correct, it disagrees with the results of 
N-body simulations: 
the PS formula overestimates the abundance of haloes near the characteristic mass 
$M_{\ast}$ and underestimates the abundance in the high mass tail (Efstathiou et al. 1988; Lacey \& Cole 1994; Tozzi \& Governato 1998; Gross et al. 1998; Governato et al. 1999). 
The J01 result is also not plotted since that mass function fits much of R03 data well at z$=$0, but diverges from R03 simulation results once well below the
limit of its empirical fit of $\ln\sigma^{-1}=-1.2$, which corresponds to $\simeq4\times10^{11} h^{-1} M_{\odot}$ with $\sigma_{8}=1.0$.
In Fig. 1, I compare the mass function of the present paper with ST, and with R03 simulation results at several
redshifts. In the figure, the solid line represents the ST mass function at $z=0, 5, 8, 15$, going from right to left, respectively. The dashed line the mass function of the present paper for the same values of the redshift, the errorbars with open squares, crosses, open triangles and solid triangles
represents R03 at $z=0, 5, 8, 15$.
Fig. 1 shows that the ST function provides a good fit to R03 data, except at very high redshifts, where it significantly overpredicts the halo abundance. At all
redshifts up to z$=$10, the difference is $\simlt 10 \%$ for each of our well sampled mass bins.  However, the ST  function begins to
overpredict the number of haloes increasingly with redshift for z$\simgt$10, up to $\sim 50$\% by z$=$15.  The simulation mass
functions appear to be generally steeper than the ST function, especially at high redshifts. This is in agreement with the theoretical mass function calculated in the present paper which gives a better description of the R03 mass function for higher values of $z$ for which the ST mass function overpredicts the simulation results. 

In Fig. 2, I plot the mass function for all of our outputs in the $f(\sigma)- \ln(\sigma^{-1})$ plane.  Large values of $\ln\sigma^{-1}$
correspond to rare haloes of high redshift and/or high  mass, while
small values of $\ln\sigma^{-1}$ describe haloes of low mass and
redshift  combinations.  
The solid line is the ST mass function while the dashed line the one obtained in the present paper and the dotted line represents 
Eq. (\ref{eq:reed}). The ST and the mass function of the present paper differs more in the high mass region, where the mass function of the present paper is steeper than ST
and in better agreement with numerical simulations data than ST mass function. 
The ST function fits the simulated mass function to better than 10$\%$ over the range of -1.7 $\leq
\ln\sigma^{-1} \leq$ 0.5 while it  
appears to significantly overpredict haloes for $\ln\sigma^{-1} \geq$ 0.5.  
The magnitude of the ST overprediction at
high values of ln$\sigma^{-1}$ is consistent with being a function
purely of ln$\sigma^{-1}$ rather than redshift, a natural consequence
of the fact that  the mass function is self similar in time (\eg
Efstathiou \etal 1988; Lacey \& Cole 1994; J01).
J01 also found an overprediction by the ST function
for $\ln\sigma^{-1} \simgt 0.75$, which with their larger simulation
volumes, corresponded primarily to objects of
z$\leq$2 and of much higher mass.  Additionally, J01 found the mass function
to be invariant with redshift within their own results.  
The empirical adjustment to the ST mass function (Eq. (\ref{eq:reed})), dotted line, describes much better numerical simulations data: for -1.7 $\leq \ln\sigma^{-1} \leq$
0.5, Eq. (\ref{eq:reed}) matches R03 data to better than 10$\%$ for well-sampled
bins, while  for 0.5 $\leq \ln\sigma^{-1} \leq$ 0.9, where Poisson
errors are larger, data is matched  to roughly 20$\%$. 
In Fig. 3, I compare the mass function of the present paper
with the simulated mass function of R03 and with the ST prediction, by plotting the residuals.
Solid straight line is the ST prediction, dashed line 
is R03 empirical adjustment to the ST function and the solid line the prediction of the present paper. Errorbars represent the R03 simulations for $0 \leq z \leq 2$ (filled squares), $2 \leq z \leq 6$ (filled triangles), $6 \leq z \leq 10$ (filled circles) and $10 \leq z \leq 15$ (open squares).
The plot also shows that the residuals for the model of the present paper (solid line) are almost coincident with those of the modified ST of R03 till when $\ln\sigma^{-1} \leq$ 0.75, and at larger values of $\ln\sigma^{-1}$, (namely $\ln\sigma^{-1} \simeq 1.2$), there is a difference of less than 10\%.
%
%
As noticed by R03, there is a large apparent scatter of the mass function for
ln$\sigma^{-1}\simgt$ 0.5 due to larger poisson errors in this
range.  
As done by R03, for large ln$\sigma^{-1}$, they estimated the uncertainty in the
mass function due to cosmic variance by estimating the contribution of
linear fluctuations  on the scale of the box size.  
The resulting estimate for the
uncertainty in the mass function due to cosmic variance is 
smaller than Poisson error limit
of 20$\%$ for Fig. 3.
However, it approaches the Poisson error limit of
20$\%$ for $z=$14.5 output. 
Thus, while cosmic variance is a significant source of
error where the mass function is steepest, it is unlikely to entirely
account for our discrepancy with the S-T function.  
As previously reported, the magnitude of the ST overprediction at
high values of ln$\sigma^{-1}$ is consistent with being a function
purely of ln$\sigma^{-1}$ rather than redshift, a natural consequence
of the fact that  the mass function is self similar in time (\eg
Efstathiou \etal 1988; Lacey \& Cole 1994; J01).
%
%

YNY and Warren et al. (2005) made a similar choice, namely they introduced an empirical mass function obtained from a fit to their simulations 
that gives a better fit to simulations than ST model.
It is important to stress that even if   
the functional forms proposed in R03, YNY and Warren et al. (2005) provide a better fit to simulations when compared
with the ST functional form, they are   
not based on theoretical background. The function obtained in the present paper, similarly, for example, to R03 provides a better fit 
to simulations than the ST functional form, and at the same time has been obtained from solid physical, theoretical, arguments.
The better agreement observed between the mass function of the present paper and R03 simulations, when compared with the ST, 
is connected to the shape of the barrier ($\delta_{\rm c}$).
The shape of the barrier given in Eq. (\ref{eq:maa}) is a direct consequence of the angular momentum acquired by the proto-structure during evolution and the effects of the cosmological constant. 
Taking account of the effects of asphericity and tidal interaction with neighbors, Del Popolo \& Gambera (1998),
showed that the threshold is mass dependent, and in particular that of the set of 
objects that collapse at the same time, the less massive ones must initially have been denser than the more massive, 
since the less massive ones would have had to hold themselves together against stronger tidal forces.  Similarly to ST, the barrier increases with $S$ (decrease with mass, $M$) differently from other models (see Monaco 1997a, b). 
The decrease of the barrier with mass means that, in order to form structure, more massive peaks must
cross a lower threshold, $\delta_c(\nu,z)$, with respect to under-dense ones.
At the same time, since the
probability to find high peaks is larger in more dense regions, 
this means that, statistically, in order to form structure, 
peaks in more dense
regions may have a lower value of the threshold, $\delta_c(\nu,z)$, with respect
to those of under-dense regions.
This is due to
the fact that less massive objects are more influenced by external tides, and
consequently they must be more overdense to collapse by a given time.
In fact, the angular momentum acquired by a shell centered on a peak
in the CDM density distribution is anti-correlated with density: high-density
peaks acquire less angular momentum than low-density peaks
(Hoffman 1986; Ryden 1988).
A larger amount of angular momentum acquired by low-density peaks
(with respect to the high-density ones)
implies that these peaks can more easily resist gravitational collapse and consequently 
it is more difficult for them to form structure.
%
%
Therefore, on small scales, where the shear is statistically greater,
structures need, on average, a higher  density contrast to collapse.

It is evident that the effect of a non-zero cosmological
constant adds to that 
of angular momentum. 
The effect of a non-zero cosmological constant is that of
slightly changing the evolution of the multiplicity function with respect to
open models with the same value of $\Omega_0$. This is caused by the fact that
in a flat universe with $\Omega_{\Lambda}>0$, the density of the universe remains close
to the critical value later in time, promoting perturbation growth
at lower redshift. The evolution is more rapid for larger values (in
absolute value) of the spectral index, $n$.

ST model was introduced at the beginning as a fit to the GIF simulations and in  a subsequent paper (SMT) was recognized the importance of aspherical collapse in the  
functional form of the mass function. The effects of asphericity were taken into account by changing the functional form of the critical overdensity (barrier) by means of a simple intuitive parameterization of elliptical collapse of isolated spheroids. The model proposed in the present paper has several similitudes with ST and ST1 models, namely it uses the excursion 
set approach as extended by ST1 to calculate the multiplicity function, but at the same time it differs from ST and ST1 for the way the barrier was calculated and for the fact that takes account of angular momentum acquisition, 
thing which is not 
taken into account into ST and ST1. These differences give rise to a multiplicity function in better agreement with simulations. 
This shows the importance of the form of the barrier.
%
The improvement of the model of the present paper and ST model with respect to Press \& Schechter is probably connected also to the fact that incorporating the non-spherical collapse with increasing barrier in the excursion set approach results in a model in which fragmentation and mergers may occur, effects important in structure formation. In the case of
non-spherical collapse with increasing barrier, a small fraction of
the mass in the Universe remains unbound, while for the spherical
dynamics, at the given time, all the mass is bound up in collapsed
objects. 
If the barrier
decreases with S (Monaco 1997a,b), this implies that all walks are
guaranteed to cross it and so there is no fragmentation associated
with this barrier shape.

In other words, the excursion set approach with a barrier taking account effects of physics of 
structure formation gives rise to good approximations to the numerical multiplicity function: the  approximation goodness increases with a more improved form of the barrier (taking account more and more physical effects: angular momentum acquisition, non zero cosmological constant, etc). 
Another important aspect of the quoted method is its noteworthy versatility: for example it is very easy to take account of the presence of a non zero 
cosmological constant englobing it in the barrier. I recall that the YNY numerical multiplicity function assumes a non zero cosmological constant while the theoretical models (ST, ST1, J01) does not take this into account. 


\begin{figure}
\centerline{\hbox{
\psfig{file=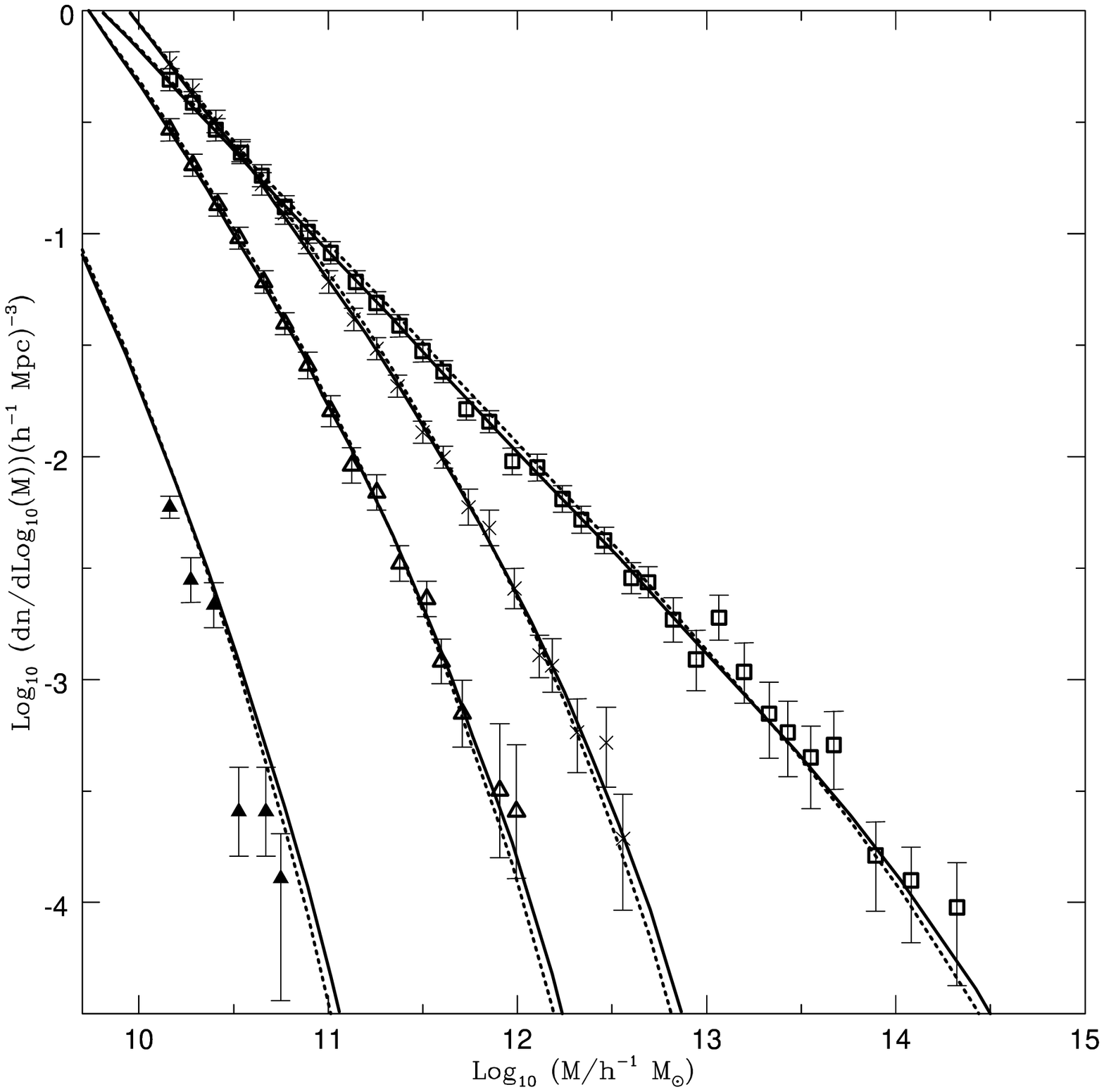,width=16cm}
}}
\caption[]{Comparison of the mass function evolution calculated in the present paper with ST mass function and R03 simulations.  
Solid curves are the Sheth \& Tormen function at z$=$0, 5, 8, \& 15 (from right to left).  
The dashed line the mass function of the present paper for the same values of the redshift, the errorbars with open squares, crosses, open triangles and solid triangles
represents R03 result at the same redshift. 
}
\end{figure}

\begin{figure}
\centerline{\hbox{
\psfig{file=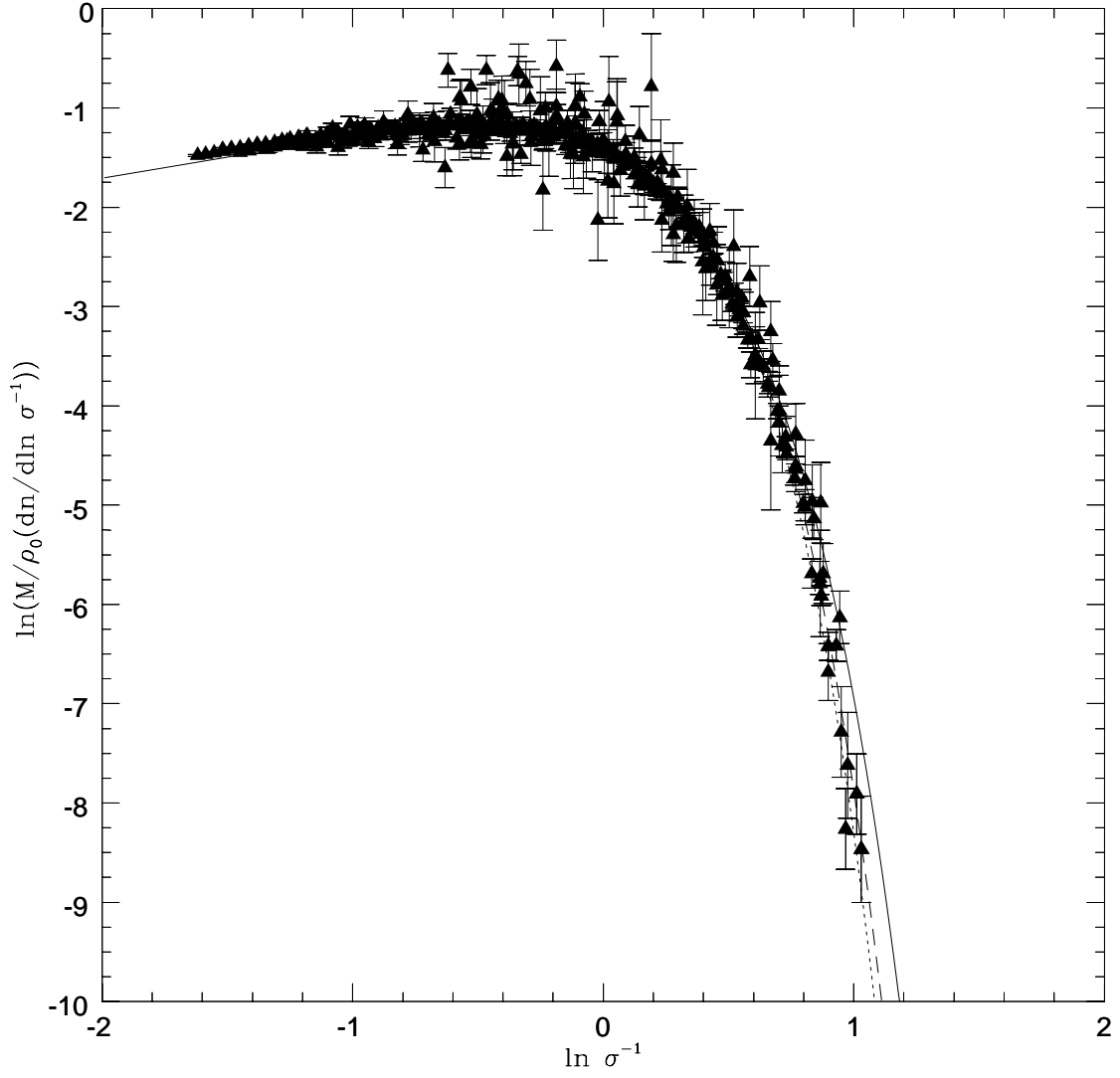,width=16cm}
}}
\caption[]{Mass function plotted in redshift independent form for all of
R03 outputs: redshifts used are 0, 1., 2., 3., 4., 5., 6.2,
7.8, 10., 12.1, 14.5. The solid line is ST prediction while the dashed and dotted line represent the result of the present paper and 
Eq. (\ref{eq:reed}), respectively.
}
\end{figure}

\begin{figure}
\centerline{\hbox{
\psfig{file=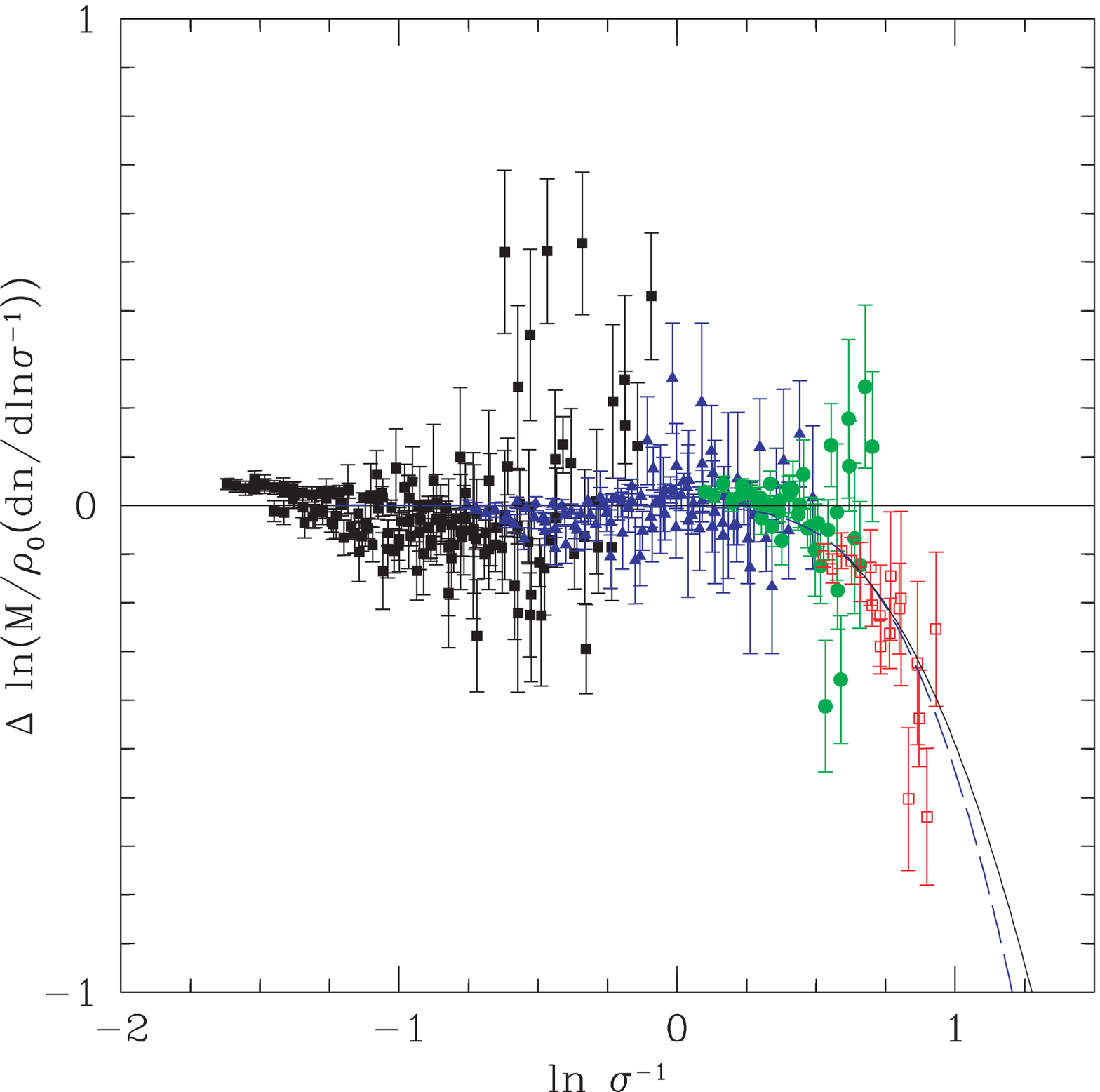,width=16cm}
}}
\caption[]{Residuals between ST prediction and the result of the present paper for the mass function of Fig. 2. Solid straight line is the ST prediction, dashed line 
is R03 empirical adjustment to the ST function and the solid line the prediction of the present paper. Errorbars represent the R03 simulations for $0 \leq z \leq 2$ (filled squares),
$2 \leq z \leq 6$ (filled triangles), $6 \leq z \leq 10$ (filled circles) and $10 \leq z \leq 15$ (open squares). 
}
\end{figure}

\section{Conclusions}

In the present paper, I compared the numerical mass function given in R03 with the theoretical mass function obtained by means of the excursion set model and an improved version of the barrier shape obtained in Del Popolo \& Gambera (1998), which implicitly takes account of tidal interactions between clusters and a non-zero cosmological constant.
I showed that the barrier obtained in the present paper
gives rise to a better description of the mass function evolution with respect to other models (ST, ST1) and that the 
agreement is based on sound theoretical models and not on fitting to simulations like in R03, YNY, or Warren et al. (2005).

The main results of the paper can be summarized as follows: \\
1) the non-constant barrier of the present paper
combined with the ST1 model gives a mass function evolution in better agreement with the N-body simulations of R03 
than other previous models (ST, ST1). \\
2) The mass function of the present paper gives a good fit to simulations results as the fit function proposed by R03, but differently from that it was obtained from a
sound theoretical background.\\
3)The excursion set model with a moving barrier is very versatile since it is very easy to introduce easily several physical effects in the 
calculation of the mass function and its evolution, just modifying the barrier.\\
4) The behavior of the mass function and its evolution at small masses is similar to that of ST, ST1, but at higher values of mass or redshift it is steeper than 
ST, ST1 in agreement with N-body simulations of R03.\\


The above considerations show that the excursion set approach that incorporates a non-spherical collapse which 
takes account of angular momentum acquisition and a non-zero cosmological constant 
gives accurate predictions for a number of statistical quantities associated with the formation and clustering of dark matter haloes.
The improvement is probably connected also to the fact that incorporating the non-spherical collapse with increasing barrier in the excursion set approach results in a model in which fragmentation and mergers may occur, effects important in structure formation.
Moreover, the effect of a non-zero cosmological
constant adds to that 
of angular momentum
%
slightly changing the evolution of the multiplicity function with respect to
open models with the same value of matter density parameter. 

\section*{Acknowledgments}
The author would like to thank D. Reed for having kindly provided his simulations data.
 
%

\end{document}